\documentclass[pra,twocolumn,aps,showpacs,preprintnumbers,amsmath,amssymb,superscriptaddress]{revtex4-1}

\usepackage{graphicx}
\usepackage{dcolumn}
\usepackage{bm}
\usepackage{amsmath}
\usepackage{cancel}
\usepackage{epstopdf}
\usepackage{tikz}

\begin{document}


\title{Nonlinear quantum error correction}

\author{Maximilian Reichert}
\affiliation{State Key Laboratory of Precision Spectroscopy, School of Physical and Material Sciences,East China Normal University, Shanghai 200062, China}
\affiliation{New York University Shanghai, 1555 Century Ave, Pudong, Shanghai 200122, China}
\affiliation{Technische Universit\"at Braunschweig,  D–38106 Braunschweig, Germany}

\author{Louis W. Tessler}
\affiliation{New York University Shanghai, 1555 Century Ave, Pudong, Shanghai 200122, China}
\affiliation{Department of Physics and Astronomy, Macquarie University, Sydney, NSW 2109, Australia}

\author{Marcel Bergmann}
\affiliation{Institute of Physics, Johannes Gutenberg-Universit\"{a}t Mainz, 55099 Mainz, Germany}

\author{Peter van Loock}
\affiliation{Institute of Physics, Johannes Gutenberg-Universit\"{a}t Mainz, 55099 Mainz, Germany}

\author{Tim Byrnes}
\email{tim.byrnes@nyu.edu}
\affiliation{New York University Shanghai, 1555 Century Ave, Pudong, Shanghai 200122, China}
\affiliation{State Key Laboratory of Precision Spectroscopy, School of Physical and Material Sciences,East China Normal University, Shanghai 200062, China}
\affiliation{NYU-ECNU Institute of Physics at NYU Shanghai, 3663 Zhongshan Road North, Shanghai 200062, China}
\affiliation{National Institute of Informatics, 2-1-2 Hitotsubashi, Chiyoda-ku, Tokyo 101-8430, Japan}
\affiliation{Department of Physics, New York University, New York, NY 10003, USA}

\date{\today}

\begin{abstract}
We introduce a theory of quantum error correction (QEC) for a subclass of states within a larger Hilbert space.  In the standard theory of QEC, the set of all encoded states is formed by an arbitrary linear combination of the codewords.  However, this can be more general than required for a given quantum protocol which may only traverse a subclass of states within the Hilbert space. Here we propose the concept of nonlinear QEC (NLQEC), where the encoded states are not necessarily a linear combination of codewords. We introduce a sufficiency criterion for NLQEC with respect to the subclass of states. The new criterion gives a more relaxed condition for the formation of a QEC code, such that under the assumption that the states are within the subclass of states, the errors are correctable. This allows us, for instance, to effectively circumvent the no-go theorems regarding optical QEC for Gaussian states and channels, for which we present explicit examples.
\end{abstract}

\maketitle

\section{Introduction}
Quantum error correction (QEC) is one of the fundamental techniques in quantum information that allows for a way to perform quantum computation in the presence of decoherence \cite{shor95,steane1996multiple,laflamme1996perfect,calderbank1996good,1997Gottesman,nielsen00,gottesman2009introduction,devitt2013quantum,RevModPhys.87.307,Brown2016}. Recently, there has been a great interest in new varieties of QEC codes incorporating topological quantum states \cite{kitaev2003,raussendorf2007topological,fowler2012surface,Landahl2011}, achieving high error thresholds.  Beyond quantum computing, QEC plays an important role in other quantum protocols such as long-distance quantum communication \cite{muralidharan16}.  
Most of the above QEC results have been developed in the context of the qubit model of quantum computation. Alternative encodings include qubit-into-qudit codes \cite{PhysRevA.86.022308} as well as the concept of encoding a qubit in an oscillator \cite{gkp01}. Such oscillator or, more generally, continuous-variable states may also represent the logical quantum information to be protected. For continuous-variable quantum information processing \cite{braunstein05,weedbrook12}, it has been shown that analogous results to qubit QEC can be realized by encoding a logical mode into many physical modes \cite{braunstein1998error,braunsteinnature98,slotine98,aoki09}.

One of the outstanding issues with continuous variables QEC is that the codes turn out to be ineffective against the most relevant errors such as those arising in 
photon loss (amplitude damping) channels. Such error channels are of Gaussian nature, mapping Gaussian states back onto Gaussian states and lacking the stochastic hierarchical behavior of the typical qubit error models. In fact, general no-go theorems have been derived to show that effective QEC for continuous variables in the presence of Gaussian errors requires non-Gaussian operations \cite{2009Niset,namiki2014gaussian}. Alternatively, non-Gaussian stochastic errors can be effectively suppressed by employing Gaussian codes and operations \cite{van2010note}. While some workarounds evading the no-go theorems using additional operations such as photon counting \cite{ralph2011quantum} and photon loss codes specifically adapted to such Gaussian errors  have been developed \cite{leghtas13,mirrahimi14,Haenggli2011,2020Noh,pvl16,li17,ofek16}, generally this has proved a formidable challenge which has made the qubit-based codes the favored approach for QEC.

In this paper, we propose as a new general framework called {\it nonlinear QEC} (NLQEC) which, as one consequence, effectively circumvents the above no-go theorems.  The primary assumption made in NLQEC is that during the execution of a particular quantum protocol, only a subclass of states in Hilbert space is visited. This is generally true in continuous-variable computing where the Hilbert space is either infinite \cite{braunstein05} or at least very large \cite{byrnes2012,byrnes2014,pyrkov14,pyrkov14b,ilookeke2018,abdelrahman2014}. We shall call this subclass of states the {\it alphabet states} for NLQEC, such that if a particular state takes this form, it is protected against errors.  An example of such a subclass of states in the quantum optics case are Gaussian states.  Gaussian states form an overcomplete set in Hilbert space, but do not include all possible states, such as Fock states or Schrodinger cat states.  In the standard theory of QEC, one generally constructs a code under the assumption that a particular type of error is present (e.g. in the 3-qubit bit flip code one assumes only bit flip errors can occur).  In NLQEC, the type of error {\it and} the class of states are restricted, which allows for a more relaxed QEC criterion.  The specific features of the alphabet states may be exploited to perform QEC.  Specifically,  we will show that a more relaxed condition for QEC can be derived due to the additional assumption on the states, where we obtain the analogue of the Knill-Laflamme QEC criterion \cite{2000knill} for NLQEC.  Our criterion will be closely related to the original QEC criterion in the sense that it also requires a similar structure of orthogonal subspaces for correctability.  There will also however be differences which arise from the use of  states within the subclass, which relax some of the restrictions that are present in the original formalism.  

\section{Alphabet states}

Let us start by first briefly reviewing the Knill-Laflamme QEC criterion  \cite{2000knill}.  Suppose quantum information is
encoded in a Hilbert space spanned by the basis states $ | \xi_i \rangle $, where $ i $ labels the codewords.   It is possible to construct a quantum error
correcting code if they satisfy
\begin{align}
\langle \xi_i | E^\dagger_n E_m | \xi_j \rangle = h_{nm} \delta_{ij}
\label{qecconditionold}
\end{align}
where $ E_n $ are the possible errors that the codewords may experience, and $ h_{nm} $ is a Hermitian matrix.   The criterion summarizes the essential
feature of QEC codes, that there exists a set of errors $ F_{n'} = \sum_n u_{nn'} E_n $, where $ u_{nn'}  $ is the matrix that diagonalizes $ h_{nm} $, which are guaranteed to be
(i)  in orthogonal subspaces for different $ F_{n'} $; and (ii) are a version of the original state that can be recoverable by a suitable recovery operation $ \cal R $. One
important aspect of the criterion
is that the matrix $ h_{nm} $ is independent of the states $  | \xi_{i} \rangle, | \xi_{j} \rangle $, which ensures  (ii).

In standard QEC theory, it is assumed that the states that are correctable will most generally be in an arbitrary superposition of the $ | \xi_i \rangle $ states.  However, as discussed above, for methods such as continuous variables involving a large Hilbert space, the class of states that are traversed are more restricted, and the above formalism does not take this into account.  Let us parametrize the subclass of states that may potentially be traversed by $ | \psi(\bm{\alpha}) \rangle $, where $ \bm{\alpha} $ are a set of
parameters to label these states.   We call $ | \psi(\bm{\alpha}) \rangle $ {\it alphabet states} as the set of states labeled by $ \bm{\alpha} $, which
are a much smaller set of states than an arbitrary linear combination of the basis states $ | \xi_i \rangle $.  This means that the states  $ | \psi(\bm{\alpha}) \rangle $ are not necessarily writable in terms of a particular subspace of Hilbert space. An example of such alphabet
states for quantum optics are Gaussian states.  Here it can 
be guaranteed that a state will always be of alphabet form due to the types of 
operations that are performed, i.e. using only beam splitters, phase shifters, and squeezers the states are always Gaussian. Alphabet states can be defined in many other ways; we shall consider several examples in Sec. \ref{sec:examples}.  


The key observation is that depending upon the particular set of states that will be traversed, particular types of errors that occur may not corrupt the quantum information. For example in classical computation, the state of the bits are always a product state, and is only ever $ |0 \rangle $ or $ | 1 \rangle $.  In this situation, a phase flip error $ \sigma^z $ on this (qu)bit is irrelevant as this gives a global phase, which does not affect the computation.  While in this case the (qu)bit is acting classically, other examples also exist where the alphabet states are explicitly quantum mechanical. 
Another example is the class of all coherent states, which are well-known to be robust in the presence of loss, because they are eigenstates of the annihilation operator \cite{sayrin2011real}. Using our theory, we show that this can be considered an elementary example of an NLQEC code.

\section{Nonlinear QEC criterion}
\subsection{Definitions}
Let us say that the set of states that we wish to protect are
\begin{align}
{\cal S} =  \{ | \psi( \bm{\alpha} ) \rangle | \, \bm{\alpha} \in {\cal A} \} 
\end{align}
where $ \bm{\alpha} $ is a potentially multidimensional set of parameters labeling the alphabet state and $ {\cal A} $ is the domain of $ \psi $. The alphabet states $| \psi ( \bm{\alpha} ) \rangle$ span a Hilbert space $\overline{\text{span}} \{ | \psi ( \bm{\alpha } \rangle \} = \mathcal{C}$ which shall be separable, implying the existence of a countable basis. Here we clarify that we do not mean separable in the sense of unentangled states, but its meaning in the context of vector spaces.  We do not consider an explicit encoder or decoder for the alphabet states as this can be performed offline, either in the state preparation or in the readout.  

The set of possible errors that may occur is
%
$ \{ E_n \} $.
%
The error-channel $\mathcal{E}$ maps the density matrix $\rho$ to the density matrix 
\begin{align}
\mathcal{E}(\rho) = \sum_{n} E_n \rho E_n^\dagger.
\end{align}
The trace is preserved, if 
\begin{align}
 \sum_{n} E_n^\dagger E_n  = I 
\end{align}
holds. Channels that do not satisfy this constraint can also be handled in our formalism. 

The existence of an NLQEC code for a given error channel $\mathcal{E}$ is defined as the existence of a recovery channel $\mathcal{R}$ consisting of the set of operators $\lbrace R_q\rbrace$ with the property $R_q E_n |\psi (\bm{\alpha}) \rangle =\lambda_{qn}(\bm{\alpha}) |\psi (\bm{\alpha}) \rangle$, where $\lambda_{qn}(\bm{\alpha})$ is a complex number. This ensures for the composite channel $\mathcal{R} (\mathcal{E}) (|\psi (\bm{\alpha})\rangle\langle \psi (\bm{\alpha})) | \sim |\psi (\bm{\alpha})\rangle\langle \psi (\bm{\alpha}))| $. Furthermore,  we require that the  recovery channel $\mathcal{R}$  happens with probability $1$.

\subsection{NLQEC criterion}
Let us define 
\begin{align}
V_{nm} (\bm{\alpha}, \bm{\beta}) \equiv \langle  \psi( \bm{\alpha}) | E_n^\dagger E_m |   \psi( \bm{\beta} ) \rangle.
\label{cmatrixdef}
\end{align}
We propose that a sufficient condition for the existence of an NLQEC for a given error channel $\mathcal{E}$ is that there exists a unitary transformation matrix $u_{n n'}$ independent of  $\bm{\alpha}, \bm{\beta}$, coefficients $c_{n}(\bm{\alpha})$ and $\Gamma_{nm}$, such that the transform of $V_{nm}( \bm{\alpha}, \bm{\beta})$ takes the form
\begin{align}
\sum_{nm} u_{n n'}^* V_{nm} (\bm{\alpha}, \bm{\beta})  u_{m m'} = c_{n'}^* (\bm{\alpha}) c_{m'}(\bm{\beta}) \langle \psi( \bm{\alpha}) |  \psi( \bm{\beta})\rangle \Gamma_{n'm'}.
\label{qeccrit}
\end{align}
The coefficients $c_n(\bm{\alpha})$ for a particular $n$ must satisfy either $|c_n (\bm{\alpha})|>0$ or $c_n(\bm{\alpha})=0$ for all $\bm{\alpha}$. Also, the matrix elements  $\Gamma_{nm}$ are symmetric and only take values $1$ and $0$ with $\Gamma_{nn}=1$, and can be written in  block diagonal form by renumbering the indices $n$. This matrix is related to the structure of the error space, which will later become apparent in the proof.

Let us first examine a few properties that follow from (\ref{qeccrit}).  In a similar way to the analysis of (\ref{qecconditionold}), we may construct  error operators
\begin{align}
F_{n'} = \sum_{n}  E_{n} u_{nn'},
\label{ftransform}
\end{align}
where $u_{nn'}$ is the unitary transformation used in \eqref{qeccrit}. The operators $F_{n'}$ represent the same quantum operation  $\mathcal{E}( \rho ) = \sum_n F_n \rho F_n^\dagger$ as the $E_n$ operators \cite{nielsen00}.
We can thus equally rewrite (\ref{cmatrixdef}) in the form
\begin{align}
\langle  \psi( \bm{\alpha} ) |  F_n^\dagger F_m  |   \psi( \bm{\beta} ) \rangle =  c_n^* (\bm{\alpha}) c_m(\bm{\beta}) \langle \psi( \bm{\alpha}) |  \psi( \bm{\beta})\rangle \Gamma_{nm} .
\label{orthogonalcrit}
\end{align}

\begin{figure}
\centering
\begin{tikzpicture}
    \node[anchor=south west,inner sep=-1] (image) at (0,0) {\includegraphics[width=\columnwidth]{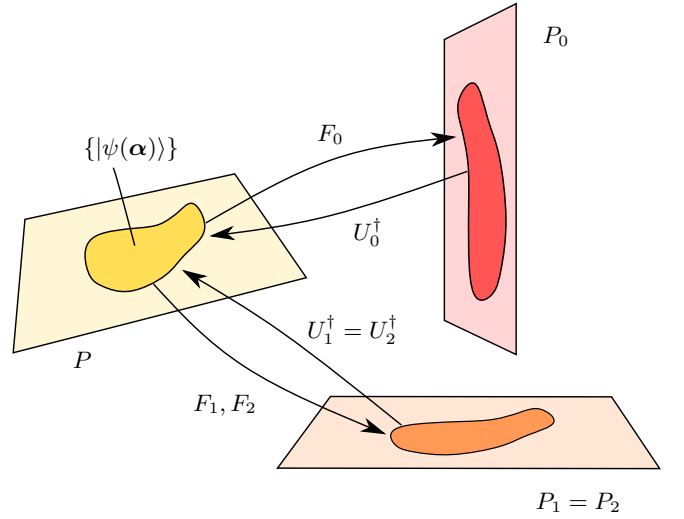}};
    \node at (7.5,0) {$P_1=P_2$};
       \node at (7.2,6.2) {$P_0$};
        \node at (4.2,4.9) {$F_0$};
         \node at (4.7,3.6) {$U^\dagger_0$};
          \node at (4.5 ,2.3) {$U^\dagger_1= U^\dagger_2$};
          \node at (2.8, 1.3) {$F_1, F_2$};
           \node at (1.55, 4.7) {$\{ | \psi (\bm{\alpha}) \rangle \}$};
           \node at (0.9, 1.9) {$P$};
\end{tikzpicture}
\caption{State structure for nonlinear quantum error correction.  The set of alphabet states $ | \psi (\bm{\alpha}) \rangle $ is depicted as the darkened area  embedded in a subspace defined by the projector $P$. For $\Gamma_{nm}=0$, the two errors $F_n$ and $F_m$ map the states to orthogonal subspaces. In the figure we have $\Gamma_{01}= \Gamma_{02}=0$. For $\Gamma_{nm}=1$, the two errors map states to the same subspace, we thus have $\Gamma_{12}=1$ for the example above. The alphabet states with errors $ F_n | \psi (\bm{\alpha}) \rangle $ again constitute a smaller set of these subspaces,  represented by the darkend areas, embedded in  larger subspaces defined by $P_n$.  Correction operations $U_n^\dagger $ return the state to its original state, completing the error correction.  In practice, an error of type $ E_n | \psi (\bm{\alpha}) \rangle $ can also be handled, which corresponds to a superposition across different spaces $ P_n $. Projection of $ E_n | \psi (\bm{\alpha}) \rangle $ by $ P_m $ collapses the state to $ F_m | \psi (\bm{\alpha}) \rangle $, after which the correction operation is identical. } \label{fig:M1}
\end{figure}

\subsection{Proof}
We prove the NLQEC criterion for a finite and linearly independent set of alphabet states $\lbrace |\psi(\bm{\alpha}_j)\rangle\rbrace$. The more technical proof for an uncountable set of alphabet states can be found in Appendix \ref{proof}. First we perform a polar decomposition of the operator product
\begin{align}
F_n P_{ \bm{\alpha}_j} &= V_n ( \bm{\alpha}_j) \sqrt{P_{ \bm{\alpha}_j} F_n^{\dagger} F_n P_{ \bm{\alpha}_j} } \nonumber \\
& = |c_n ( \bm{\alpha}_j)| V_n ( \bm{\alpha}_j) P_{ \bm{\alpha}_j} \nonumber \\
&= c_n ( \bm{\alpha}_j) U_n ({\bm{\alpha}_j} ) P_{ \bm{\alpha}_j} ,
\label{polardecomp}
\end{align}
where $U_n (\bm{\alpha}_j) =e^{-i \text{arg} (\bm{\alpha}_j)} V_n (\bm{\alpha}_j)$ and $P_{\bm{\alpha}_j} = |\psi(\bm{\alpha}_j) \rangle \langle \psi (\bm{\alpha}_j)|$ is a one dimensional projection. According to the polar decomposition theorem, $U_n (\bm{\alpha}_j)$ is a isometry that generally is not unique. In the following, we will show that this freedom of choice enables us to find an isometry $U_n$ that is independent of $\bm{\alpha}_j$. This $\bm{\alpha}$-independence ensures that we can recover the state without  knowledge of the error that has occurred. We explicitly construct this operator by defining its action on states $| \phi \rangle = \sum_{\bm{\alpha}_j} a_j  | \psi( \bm{\alpha}_j) \rangle \in \text{span} \lbrace |\psi(\bm{\alpha}_j)\rangle \rbrace = \mathcal{C}$
\begin{align}
U_n | \phi \rangle = \sum_{\alpha_j} a_j U_n ( \bm{\alpha}_j)  | \psi( \bm{\alpha}_j) \rangle .
\label{unitarydef}
\end{align}
By definition, this operator is linear. It is also well-defined because $\lbrace  |\psi(\bm{\alpha}_j)\rangle \rbrace$ is a linearly independent set. It now follows directly that $U_n (\bm{\alpha}_j)$ can be replaced by $U_n$

\begin{align}
F_n  | \psi( \bm{\alpha}_j) \rangle = c_n ( \bm{\alpha}_j) U_n ( \bm{\alpha}_j)  | \psi( \bm{\alpha}_j) \rangle = c_{n} (\bm{\alpha}_j) U_n | \psi( \bm{\alpha}_j) \rangle  .
\label{errorisunitary}
\end{align}

We define $P$ as the projector of $\mathcal{C}$. By using \eqref{errorisunitary} and \eqref{orthogonalcrit} we find
\begin{equation}
P U_n^\dagger U_m P= P \Gamma_{nm} .
\end{equation}
Thus, $U_n$ is an isometry restricted on $\mathcal{C}$. In finite dimensions, this isometry can always be extended to a unitary operator with $U_n^\dagger U_n = I$. 

The image of $U_n$ is a subspace with the corresponding projection operator
\begin{align}
P_n = U_n P U_n^\dagger ,
\end{align}
with $P_n P_m = P_n \Gamma_{nm}$. From $\Vert (U_n-U_m)|\phi\rangle \Vert^2= 2 -2 \Gamma_{nm}$ for all $|\phi\rangle \in \mathcal{C}$, it follows that the operators $U_n$ and $U_m$ restricted to $\mathcal{C}$ are equal and map to the same subspace for $\Gamma_{nm}=1$. Because we can extent both $U_n$ and $U_m$ in the same way to unitary operators, we further find $U_n=U_m$ for the whole Hilbert space. For $\Gamma_{nm}=0$ the operators map to orthogonal subspaces.
For each distinct subspace we define a recovery operator
\begin{align}
 R_q =  U_q^{\dagger} P_q 
\label{recovery}
\end{align}
which satisfies
\begin{align}
R_q E_n | \psi (\bm{\alpha}) \rangle & = \sum_{n'} u_{n n'}^* R_q F_{n'} | \psi (\bm{\alpha}) \rangle \nonumber\\
& = \underbrace{\sum_{n'} u_{n n'}^*  c_{n'} (\bm{\alpha}) \Gamma_{qn'}}_{\equiv \lambda_{nq}(\bm{\alpha})} | \psi (\bm{\alpha}) \rangle \label{RE}
\end{align}
as is required for NLQEC recovery operators. By adding $R_0 = P$ to $\{R_q\}$ if necessary, we find, that the recovery operation is trace-preserving because of $\sum_n R_n^\dagger R_n = \sum_n P_n = I$. 

For a mixed alphabet state $\rho = \sum_j p_j  |\psi(\bm{\alpha}_j)\rangle \langle \psi (\bm{\alpha}_j)|$ the recovery yields
\begin{equation}
\mathcal{R} ( \mathcal{E} (\rho) ) =  \sum_{\bm{\alpha}_j} c(\bm{\alpha}_j) p_j | \psi(\bm{\alpha}_j)\rangle \langle \psi(\bm{\alpha}_j )|
\end{equation}
with $c(\bm{\alpha}_j) = \sum_{qn}|\lambda_{qn}(\bm{\alpha}_j)|^2$. For trace-preserving error channels we have $c(\bm{\alpha}_j)=1$ and thus a perfect recovery of the mixed state. This follows directly from  the fact, that the composite channel $\mathcal{R} (\mathcal{E} (\rho))$ is trace-preserving and linear. $\blacksquare$

\subsection{Consistency and comparison with standard QEC criterion}
We verify that (\ref{qecconditionold}) satisfies the NLQEC criterion as a particular case.  Our criterion should be a relaxed version of the standard QEC criterion, hence any QEC code that satisfies (\ref{qecconditionold}) should also satisfy (\ref{qeccrit}).  In this case our alphabet states are simply the usual set of states that are a linear combination of the codewords
\begin{align}
| \psi (\bm{\alpha}) \rangle = \sum_j \alpha_j | \xi_j \rangle .
\end{align}
Straightforward evaluation of (\ref{qeccrit}) using (\ref{qecconditionold}) yields
\begin{align}
\langle  \psi( \bm{\alpha} ) | E_n^\dagger E_m |   \psi( \bm{\beta} ) \rangle = h_{nm} \langle \psi( \bm{\alpha}) | \psi (\bm{\beta}) \rangle .
\end{align}
%
The sufficiency condition \eqref{orthogonalcrit} is also satisfied, since  the matrix $h_{nm}$ is hermitian and thus can be diagonalized as
\begin{align}
\sum_{nm} u_{n n'}^* V_{nm} (\bm{\alpha}, \bm{\beta})  u_{m m'} = c_n^* c_m \langle \psi( \bm{\alpha}) |  \psi( \bm{\beta})\rangle \delta_{nm} .
\end{align}
The error operators of the Knill-Laflamme criterion constructed with $u_{nn'}$ can be written as $F_n P = c_n U_n P$, where $P$ is the projector of the space spanned by the codewords $\lbrace |\xi_j\rangle\rbrace$. Thus, the protected states are stretched by a factor $c_n$ and unitarily transformed by $U_n$, which map to orthogonal subspaces. 

In NLQEC, the error operators \eqref{ftransform} can be interpreted in the same way, the only difference is that the stretching factor now depends on the alphabet state $|\psi(\bm{\alpha}) \rangle$. By introducing the operator $J_n$  that has the alphabet states as its eigenvectors with eigenvalues $c_n(\bm{\alpha})$, we can write $F_n P = U_n J_n P$, which follows from \eqref{errorisunitary}, where $P$ now is the projector of the space spanned by the alphabet states. In the case of no $\bm{\alpha}$-dependence, we have $J_n = c_n P$ and the error representation reduces to the Knill-Laflamme case. It is also becomes apparent, that superpositions of alphabet states generally are not recoverable, because the individual components get stretched by different amounts, thus deforming the state.

\section{Examples}
\label{sec:examples}

\subsection{Example 1: Coherent states}
To show the theory in action, we now present some examples of error channels and protected states, that satisfy NLQEC, but not the standard theory of QEC. Consider the simplified error channel modelling the loss of one photon
\begin{align}
{\cal E} & =\{  I, a \}  
\label{coherentstateerror}
\end{align}
where $a$ is the bosonic annhilation operator of the Fock space.
The protected alphabet states have to be eigenstates of the $J_n$ operators which we introduced in the previous section. In this example we have $J_0 =  I$ and $J_1 = a$. The mutual eigenstates of $J_0$ and $J_1$ are the coherent states defined via
\begin{equation}
| \psi (\alpha) \rangle = | \alpha \rangle = e^{-|\alpha|^2/2} e^{\alpha a^\dagger} | 0 \rangle = D(\alpha) | 0 \rangle
\end{equation}
where $ \alpha $ is an arbitrary complex number.
Evaluating \eqref{cmatrixdef}, we find
\begin{align}
V (\alpha, \beta) = \begin{pmatrix}
1  & \beta\\ 
\alpha^*  &  \alpha^* \beta
\end{pmatrix} \langle \alpha   | \beta \rangle  .
\end{align}
This satisfies the NLQEC criterion with $ c_0 (\alpha) = 1, c_1 (\alpha) = \alpha $ and $ \Gamma_{nm}=1$.   This implies that there is only one subspace for the two types of errors, hence there is only one correction operation. The correction operation is achieved in this case by $ R = I$.  
The intuition for this result is that for coherent states, the errors (\ref{coherentstateerror}) do not affect the alphabet states, up to a overall constant.  Since superposition of alphabet states are never generated, the constant does not affect the state, and hence the quantum information is preserved.  The corresponding case for a more realistic photon loss channel is given in Appendix \ref{sec:coherentstates}.

\subsection{Example 2: Collective dephasing}
Another simple example in the case of qubits involves a collective dephasing error channel
\begin{align}
{\cal E}= \{ F_0, F_1 \} = \{ \sqrt{p} I, \sqrt{(1-p)} Z_1 Z_2 \},
\label{error2qubit}
\end{align}
where the probability of the phase flip is $1- p $ and $ X_i, Z_i $ are the Pauli operators on the $ i $th qubit. Such a model of dephasing is considered often in the context of decoherence-free subspaces \cite{lidar2012}. For a general two-qubit state neither the NLQEC nor the standard QEC conditions are satisfied; however, with additional constraints we may satisfy the NLQEC criterion. We first consider alphabet states of the form
\begin{align}
|\psi (j,\theta,\phi) \rangle = (X_2)^j (\cos \theta | 00 \rangle + e^{i \phi} \sin \theta | 11 \rangle)
\label{psi1alphabet}
\end{align}
where $\bm{\alpha} = \{ j,\theta, \phi \}$ with $j\in \{0,1\} $ is the set of parameters that parametrizes the alphabet states. For fixed $ j $, these correspond to decoherence-free subspaces for the error channel. Evaluating \eqref{cmatrixdef} we find that 
\begin{align}
V= & \left( 
\begin{array}{ll}
p   & \sqrt{p(1-p)} (-1)^j \\
 \sqrt{p(1-p)} (-1)^{j'} & (1-p) (-1)^{j + j'}
\end{array}
\right) \nonumber \\
& \times \langle \psi(j', \theta', \phi' ) | \psi(j, \theta, \phi) \rangle .  
\label{collectivevnm}
\end{align}
This satisfies the NLQEC criterion taking
\begin{align}
c_0 (j, \theta, \phi)  & = \sqrt{p} \nonumber \\
c_1  (j, \theta, \phi) & = (-1)^j \sqrt{1-p} \label{csolcollective}
\end{align}
and $ \Gamma_{nm} = 1 $.   In this case the recovery operator is $ R = I $, since the error operator only creates a global phase on the state.  The above is an exact example of a nonlinear QEC that does not satisfy the Knill-Laflamme QEC criterion, since $(\ref{csolcollective})$ depends on the alphabet parameter $j$ and so the global phase can become a relative one.

Another way to satisfy the NLQEC conditions is to take the alphabet states as
\begin{align}
| \psi (\theta) \rangle = & \frac{1}{\sqrt{2}} \Big( e^{i \phi_0} \cos \theta | 00 \rangle +\sin \theta | 01 \rangle \nonumber \\
& + \cos \theta | 10 \rangle
+ e^{i \phi_0 } \sin \theta | 11 \rangle \Big)
\label{psi3alpha}
\end{align}
where $ \theta $ is an alphabet parameter, but $ \phi_0 $ must be held fixed. In this case \eqref{cmatrixdef} is
\begin{align}
V = & \left( 
\begin{array}{ll}
p   & 0 \\
 0 & 1-p 
\end{array}
\right) \langle \psi(\theta' ) | \psi( \theta ) \rangle .  
\label{collectivevnm2}
\end{align}
In this case the NLQEC criterion is satisfied with $ c_0(\theta) = \sqrt{p} $, $ c_1(\theta) = \sqrt{1-p} $, and  $ \Gamma_{nm} = \delta_{nm} $.  For this example we have two subspaces, corresponding to the code space and the error space.   In this case, a $ Z_1 Z_2 $  error causes the alphabet states (\ref{psi3alpha}) to be transformed into an orthogonal state, which allows the error to be detected. The original state can be recovered by applying $ R = Z_1 Z_2 $.

\subsection{Example 3: Squeezed coherent states}
Next we consider squeezed coherent states as alphabet states 
\begin{align}
| \psi (\bm{\alpha}) \rangle = | \alpha  \rangle_\xi  =
e^{\frac{1}{2} (\xi^* a^2 - \xi a^{\dagger 2})} | \alpha \rangle = S(\xi) D(\alpha)|0\rangle ,
\end{align}
where $\xi = r e^{i\theta}$ is a complex number.
Such states are fundamental to continuous variables quantum information protocols and it is of particular interest whether NLQEC criterion can be applied.  Note that here we consider $ \alpha$ to be the parameter labeling the alphabet states, and $\xi$ is considered fixed.
Evaluating \eqref{cmatrixdef} for the simplified photon loss channel (\ref{coherentstateerror})  we find
\begin{align}
V (\alpha, \beta) = \begin{pmatrix}
1  & \Omega(\alpha,\beta)  \\ 
\Omega^*(\beta ,\alpha)  & \Omega^*(\beta ,\alpha) \Omega(\alpha,\beta) + \sinh^2 r 
\end{pmatrix} \langle \alpha   | \beta \rangle  
\label{sqcohv}
\end{align}
where 
\begin{align}
\Omega(\alpha ,\beta) = \beta \cosh r - \alpha^* e^{i \theta} \sinh r .
\end{align}
Eq. (\ref{sqcohv}) does not satisfy the NLQEC criterion for arbitrary $ \alpha,\beta$ because: (i)   the presence of the $ \sinh^2 r$ term, and (ii) the mixed $ \alpha,\beta$ dependence of the $ \Omega(\alpha ,\beta) $ function  means that the matrix cannot be written in terms of a product of two factors $ c_n^*( \alpha) c_m( \beta)  $. 

By making additional restrictions on the alphabet labels $ \alpha$, it is possible to make (\ref{sqcohv}) approximately satisfy the NLQEC criterion.  Regarding (i), the $ \sinh^2 r$ is negligible if  $|\alpha|,|\beta |\gg 1$.  For (ii), to see why the mixed dependence on $\alpha, \beta $ can be removed, we rewrite 
\begin{equation}
\Omega (\alpha,\beta) \langle \alpha | \beta\rangle = \Omega (\alpha,\alpha)\langle \alpha|\beta\rangle + \epsilon (\beta,\alpha)
\end{equation}
with $\epsilon(\beta,\alpha) = (\beta^* - \alpha^*)e^{i\theta} \sinh r \langle \alpha|\beta \rangle$.
In the case $|\beta - \alpha| \approx 1$, we have $|\epsilon (\beta,\alpha)| \approx e^{-1/2} \sinh r$ and because $|\alpha|,|\beta|\gg 1$, the $\epsilon (\beta,\alpha)$ is small in comparison to $\Omega (\beta,\alpha)\langle \alpha |\beta \rangle$ and can be neglected. In the cases  $|\alpha - \beta| \ll1$ and  $|\alpha - \beta| \gg 1$, we find $\epsilon (\beta,\alpha) \approx 0$ and  we may write
\begin{align}
V (\alpha, \beta ) \approx \begin{pmatrix}
1  & \Omega(\beta,\beta )   \\ 
\Omega^*(\alpha ,\alpha )  & \Omega(\alpha,\alpha) \Omega^*(\beta ,\beta ) 
\end{pmatrix}  \langle \alpha | \beta \rangle
\label{sqcohv2}
\end{align}
which satisfies the NLQEC criterion approximately.  Again, in this case there is only one error space because of $\Gamma_{nm}=1$. As the recovery operator we take again $R=I$ for both types of errors.

It is interesting to see why the additional restriction allows us to satisfy the NLQEC criterion more from physical grounds.  Unlike coherent states which are exact eigenstates of the annihilation operator, a photon loss event on a squeezed coherent state produces an orthogonal component defined as
\begin{align}
| \perp_\alpha  \rangle = a | \alpha \rangle_{\xi} - \alpha   | \alpha \rangle_{\xi} .
\end{align}
We need to ensure that after an error occurs, the state is  not disturbed so much that the orthogonal component becomes the dominant contribution.  We thus demand that
\begin{align}
    \frac{ \langle \perp_\alpha | \perp_\alpha \rangle}{{_{\xi}} \langle \alpha | a^\dagger a | \alpha \rangle_{\xi}  } & = \frac{\sinh^2 r }{ \Omega(\alpha,\alpha ) \Omega^*(\alpha ,\alpha ) +\sinh^2 r }   
\end{align}
is small.  This can be achieved if $ | \alpha | \gg \sinh r$, which recovers  our previous requirement. This restriction avoids inclusion of the squeezed vacuum which contains only even Fock states, which are highly susceptible to the state changing effects under photon loss. 

\subsection{Example 4: Cat states}
The previous example only involved one subspace.  A simple two subspace example for optical states can be constructed using cat states that are defined as
\begin{align}
 | \alpha_e \rangle =  \frac{1}{\sqrt{2 (1+e^{-2|\alpha|^2})}}\left(| \alpha \rangle + | - \alpha \rangle \right) .
 \label{evencats}
\end{align}
The alphabet of states is defined for $ \alpha $ in  half of the complex plane away from the line $ \text{Re} (\alpha) =0 $, for example  $\text{Re} (\alpha) \gg 0$.  Considering again the simplified photon loss channel  $ {\cal E}  =\{ I, a \}  $, we obtain
\begin{align}
V(\alpha, \beta) \approx \begin{pmatrix}
\langle \alpha_e | \beta_e \rangle & 0 \\ 
0 &  \alpha^* \beta \langle \alpha_o | \beta_o \rangle
\end{pmatrix} \label{innerproducts}
\end{align}
where $| \alpha_o \rangle =  \frac{1}{\sqrt{2}}\left(| \alpha \rangle - | - \alpha \rangle \right)$ is an odd cat state. Because of $\text{Re} (\alpha) \gg 0$,  the normalisation constant can be approximated as $1/\sqrt{2}$ for both types of cat states. The two error spaces are orthogonal to each other which is implied by $\Gamma_{nm} = \delta_{nm}$. Let us compare the two inner products in \eqref{innerproducts}, 
\begin{align}
\langle \alpha_e | \beta_e \rangle -  \langle \alpha_o | \beta_o \rangle =\langle -\alpha | \beta \rangle + \langle \alpha | - \beta \rangle .
\end{align}
This is small for our choice of alphabet states $  | \alpha_e \rangle $.  
The NLQEC criterion is approximately satisfied because of $\langle \alpha_e | \beta_e \rangle \approx \langle \alpha_o | \beta_o \rangle$. The main findings remain true for the more realistic photon loss channel (see Appendix \ref{sec:coherentstates}). 

As the NLQEC theory suggests, the error operator $a$ can be decomposed as $ a = T \sqrt{\hat{n}}$, where $ J =\sqrt{\hat{n}}$ is the square root of the photon number operator and $U = T$ is the left shift operator.
For $|\alpha | \gg 0$, the cat states are indeed approximate eigenstate of $J$ and as in the previous example we consider them as alphabet states in the approximate sense. We find
\begin{equation}
J | \alpha_e \rangle = \sqrt{\hat{n}} | \alpha_e \rangle \approx |\alpha| | \alpha_e \rangle 
\end{equation}
where the symbol $\approx$ means that the two states are close to each other in Hilbert space.
Because here $\Gamma_{nm} = \delta_{nm}$ we have two orthogonal error spaces $P_0, P_1$ with even and odd Fock number states.  If the state is found to be in the even Fock space, the recovery operator $R_0 = I$ is applied, which recovers the state perfectly. For the odd subspace, one applies $R_1 = T^\dagger$ which recovers the state approximately $ T^\dagger a | \alpha_e \rangle = T^\dagger T \sqrt{\hat{n}} | \alpha_e \rangle  $, where $ T^\dagger T = I - |0\rangle \langle0|$. The fidelity, which is a measure of the quality of the recovery is given by the square of the overlap of the recovered state and the original state. We find for the fidelity
$
\frac{1}{4|\alpha|^2} (\langle \alpha | \sqrt{\hat{n}} |\alpha\rangle + \langle - \alpha | \sqrt{\hat{n}} |-\alpha\rangle )^2 \approx 1 - \frac{1}{4 |\alpha|^2}
$
where the value $1$ corresponds to a perfect recovery.

To understand from physical grounds why the above state satisfies the NLQEC criterion, consider a single photon loss event on the even cat state $  | \alpha_e \rangle $.  This produces the state 
\begin{align}
a | \alpha_e \rangle = \alpha | \alpha_o \rangle, 
\end{align}
which is simply an odd cat state up to a constant factor.  Since the odd cat states are in a different subspace to the even cat states, this error event is detectable, and also correctable.  This would not be true if the cat state was not defined as $  | \alpha_e \rangle $, for example with an asymmetric amplitude of the cat superposition.  In this case a photon loss event would corrupt the nature of the state, and does not satisfy the NLQEC criterion.

\section{Summary and conclusions}

We have introduced a theory of NLQEC, where only a subclass of states is considered to be traversed during a particular quantum algorithm. The subclass of states is not necessarily defined by a linear superposition of codewords, and is nonlinear in this sense.  Our main result is the sufficiency condition (\ref{qeccrit}) for the existence of a QEC code for alphabet states.  This is a more relaxed condition than the standard QEC criteria in the sense that any QEC code that satisfies (\ref{qecconditionold}) also satisfies (\ref{qeccrit}), but the reverse is not true.  The key feature is that it takes into account of the limited class of states that may be physically realized for particular quantum computing platforms.  The best-known example of this are Gaussian optical states in continuous variable quantum information \cite{braunstein05}, but also could be equally applicable for other quantum computing approaches \cite{byrnes2012,byrnes2014}.  Some simple examples such as coherent states, two-qubit states, and  squeezed coherent states were shown to be either exact or approximate NLQEC codes.  This formalizes the well-known notion that coherent states are robust in the presence of loss.  The parameter ranges where the alphabet states satisfy the NLQEC criterion coincide with physical arguments based on the effect of errors applied on alphabet states.  

Due to the additional assumption on the nature of the states that are traversed during the quantum protocol,  well-known no-go theorems concerning QEC with Gaussian states, operations, and channels, do not apply to NLQEC.  Since the types of operations that are needed for a particular implementation of quantum information processing are usually known in advance, the additional assumption does not constitute any additional practical restriction. For example, for optical Gaussian alphabet states, constructing a protocol with beam splitters, phase shifters, and squeezers guarantee that the states are Gaussian.  
We note that even with the NLQEC theory, it is not possible to argue that all Gaussian states are robust in the presence of loss. However, as shown with the squeezed coherent state example, additional constraints to the alphabet state can be introduced to satisfy the NLQEC code approximately \cite{Schumacher,leung97,hayden}.  It is interesting that the additional constraints that must be added can be explained rather intuitively from physical grounds, despite the more abstract origins of the theory.  This may give the opportunity for the NLQEC criterion to be used as a guide to construct codes that could be tailored to the naturally present errors, such that the alphabet states are robust by construction.


\section*{Acknowledgments}
We thank Barry Sanders for discussions on related topics.  TB is supported by the National Natural Science Foundation of China (62071301); State Council of the People’s Republic of China (D1210036A); NSFC Research Fund for International Young Scientists (11850410426); NYU-ECNU Institute of Physics at NYU Shanghai; the Science and Technology Commission of Shanghai Municipality (19XD1423000); the China Science and Technology Exchange Center (NGA-16-001); the NYU Shanghai Boost Fund.  PvL acknowledges support from the BMBF in Germany via Q.Link.X.

\appendix

\section{Proof of the NLQEC criterion} 
\label{proof}

Let us assume the NLQEC criterion is satisfied. Thus, there exists a unitary matrix $u_{nn'}$ such that for a given set of errors and alphabet states
\begin{align}
\sum_{nm} u_{n n'}^* V_{nm} (\bm{\alpha}, \bm{\beta})  u_{m m'} = c_{n'}^* (\bm{\alpha}) c_{m'}(\bm{\beta}) \langle \psi( \bm{\alpha}) |  \psi( \bm{\beta})\rangle \Gamma_{n'm'}.
\label{qeccritapp}
\end{align}
is satisfied. As described in the main text, we may equivalently write this as
\begin{align}
\langle  \psi( \bm{\alpha} ) |  F_n^\dagger F_m  |   \psi( \bm{\beta} ) \rangle =  c_n^* (\bm{\alpha}) c_m(\bm{\beta}) \langle \psi( \bm{\alpha}) |  \psi( \bm{\beta})\rangle \Gamma_{nm} .
\label{orthogonalcritapp}
\end{align}
We now prove that \eqref{qeccritapp} is a sufficient condition for the existence of an NLQEC code. For that, we will show that the error operators $F_n$ act on all of the alphabet states as isometries up to a scaling, that is $F_n | \psi (\bm{\alpha}) \rangle = c_n (\bm{\alpha}) U_n |\psi (\bm{\alpha}) \rangle$. We explicitly construct the operator $U_n$ by introducing a basis of $\mathcal{C}$ and prescribe how the operator acts on these basis states. We then show that two choices of the error operators $ F_n $ and $F_m$  give a mapping to orthogonal subspaces if $\Gamma_{nm}=0$; if on the other hand $\Gamma_{nm}=1$, they map to the same error space as can be seen in Fig. 1 in the main text. This orthogonal structure ensures that an error measurement reveals the error that occurred to an alphabet state without destroying the alphabet state itself. The operator $U_n^\dagger$ is then  used to define the operator $R_n$ which recovers the state.

Let us start the proof. We assume that the set of alphabet states is uncountable. The proof for a countable set of alphabet states is analogous.  For convenience, we will work with the error operators
\begin{align}
F_{n'} = \sum_{n}  E_{n} u_{nn'}  .
\label{ftransformapp}
\end{align}
First we examine how they act on a single alphabet state. For that, we introduce the one dimensional projection operator  $P_{\bm{\alpha}}=  | \psi( \bm{\alpha}) \rangle \langle \psi( \bm{\alpha} )|$  and do a polar decomposition 
\begin{align}
F_n P_{ \bm{\alpha}} &= V_n ( \bm{\alpha}) \sqrt{P_{ \bm{\alpha}} F_n^{\dagger} F_n P_{ \bm{\alpha}} } \nonumber \\
& = |c_n ( \bm{\alpha})| V_n ( \bm{\alpha}) P_{ \bm{\alpha}} \nonumber \\
&= c_n ( \bm{\alpha}) U_n ({\bm{\alpha}} ) P_{ \bm{\alpha}} 
\label{polardecompapp}
\end{align}
where $e^{i \text{arg} ( c_n ( \bm{\alpha}))} U_n ({\bm{\alpha}}) = V_n ( \bm{\alpha})$. According to the polar decomposition theorem, $U_n ( \bm{\alpha})$ is a partial isometry which is not uniquely determined. As the name implies, partial isometries are isometries on a subspace of the Hilbert space. They give 0 on the orthogonal complement of that subspace.  We generally have $U_n^\dagger U_n \neq U_n U_n^\dagger$. In the following we will show, that the partial isometry in \eqref{polardecompapp} can be chosen independently of $\bm{\alpha}$, that is $U_n ( \bm{\alpha}) =U_n$. This $\bm{\alpha}$-independence ensures that a recovery $U_n^\dagger$ can be performed without knowledge of the alphabet state $| \psi (\bm{\alpha}) \rangle$.

To construct the operator $U_n$, we first define a basis of $\mathcal{C}$ and specify how $U_n$ acts on the basis states.  As we show in Sec. \ref{app:basis}, there exists a countable and possibly overcomplete basis $\{ |\psi (\bm{\alpha}_j) \rangle \}$ with the property, that every finite subset of the basis is linearly independent. This property is important for the construction of $U_n$.  We define $U_n$ by  $U_n |\psi (\bm{\alpha}_j)\rangle = U_n (\bm{\alpha}_j) |\psi (\bm{\alpha}_j) \rangle$.  The action of $U_n$ on an arbitrary state $| \phi \rangle = \sum_{j}^{\infty} a_j | \psi (\bm{\alpha}_j) \rangle$ in $\mathcal{C}$ shall then be given by
\begin{align}
U_n | \phi \rangle = \sum_{\bm{\alpha}_j}^{\infty} a_j U_n ( \bm{\alpha}_j)  | \psi( \bm{\alpha}_j) \rangle .
\label{unitarydefapp}
\end{align}
The proof that this operator is well-defined can be found in Sec.  \ref{welldefined}.

 Another basis $\{ |\psi (\tilde{\bm{\alpha}}_j) \rangle \}$, that satisfies the required condition of linear independence for finite subsets, gives rise to the operator $\tilde{U}_n$. In Sec. \ref{app:independence}, we show that the constructed operator is independent of its defining bases, that is, all basis give rise to the same operator  $\tilde{U}_n |\phi\rangle = U_n|\phi\rangle$.

Now we finally can show that the non-uniquely determined $U_{n}(\bm{\alpha})$ in \eqref{polardecompapp} can be chosen to be independent of $\bm{\alpha}$. For every  $\bm{\alpha}$ we can always find a basis $\{ |\psi (\bm{\alpha}_j) \rangle \}$, such that $| \psi (\bm{\alpha}) \rangle \in \{ |\psi (\bm{\alpha}_j) \rangle \}$, that has linearly independent finite subsets. By definition we have $ U_n|\psi(\bm{\alpha}) \rangle =U_n (\bm{\alpha})|\psi(\bm{\alpha}) \rangle $ \eqref{unitarydefapp}. Since all bases give rise to the same operator $U_n$, we find 
\begin{align}
F_n  | \psi( \bm{\alpha}) \rangle = c_n ( \bm{\alpha}) U_n ( \bm{\alpha})  | \psi( \bm{\alpha}) \rangle = c_{n} (\bm{\alpha}) U_n | \psi( \bm{\alpha}) \rangle 
\label{errorisunitaryapp}
\end{align}
for all $\bm{\alpha}$.  This proves, that $F_n$ transforms all of the alphabet states as an partial isometry up to a scaling $c_n (\bm{\alpha})$. We can rewrite the above equation as
\begin{equation} \label{FUJ}
F_n P = U_n J_n P
\end{equation}
where $J_n$ has the alphabet states $|\psi (\bm{\alpha})\rangle$ as its eigenstates with eigenvalues $c_n (\bm{\alpha})$. This representation of $F_n$ demonstrates, that superpositions of alphabet states generally get deformed because they are not an eigenstate of $J_n$ and are thereby not recoverable.

For unambiguous error detection, the error spaces defined by $F_n$ and $F_m$ either have to be orthogonal or the same. The projection operators of the error spaces are defined by
\begin{align}
P_n = U_n P U_n^\dagger .
\end{align}
With $PU_n^\dagger U_m P = P \Gamma_{nm}$ from \eqref{unitaryproof}  follows
\begin{align}
P_n P_m = P_n \Gamma_{nm} .
\end{align}
Thus, for $\Gamma_{nm}=0$ the error spaces are indeed orthogonal, whereas alphabet states are being sent to the same error space for $\Gamma_{nm}=1$.

For each subspace we now define the recovery operator
\begin{align}
R_q = P U_q^{\dagger} P_q
\label{recoveryapp}
\end{align}
consisting of a syndrome measurement $P_q$ which reveals the error that occurred and a subsequent recovery $PU_q^\dagger$.

The recovery operators satisfy
\begin{align}
R_q E_m | \psi (\bm{\alpha}) \rangle & = \sum_{m'} u_{m' m}^* R_q F_{m'} | \psi (\bm{\alpha}) \rangle \nonumber\\
& = \underbrace{\sum_{m'} u_{q m}^*  c_{m'} (\bm{\alpha}) \Gamma_{qm'}}_{\equiv \lambda_{qm}(\bm{\alpha})} | \psi (\bm{\alpha}) \rangle \label{REapp} ,
\end{align}
which is required for NLQEC recovery, where we used $P U_q^\dagger U_m P = P \Gamma_{qm}$ to get from the first to the second line. Finally, we define the recovery operation as $\mathcal{R} ( \rho) = \sum_{n} R_n \rho  R_n^\dagger$. Using \eqref{REapp}, we find
\begin{align}
\mathcal{R}(\mathcal{E}( | \psi (\bm{\alpha} )\rangle \langle \psi (\bm{\alpha})| )) = c(\bm{\alpha})  | \psi (\bm{\alpha}) \rangle \langle \psi (\bm{\alpha})| \label{fullrecov}
\end{align}
where $c(\bm{\alpha}) = 1$ for trace-preserving channels. This completes the proof of the criterion \eqref{orthogonalcritapp}.

The recovery operators do not necessarily satisfy $\sum_k R_k^\dagger R_k = I$ in an infinite dimensional Hilbert space. However, by construction, the recovery operation happens with probability 1, which follows from  $\text{tr} (\mathcal{R} (\mathcal{E} (| \phi \rangle \langle \phi |)) = \text{tr} (\mathcal{E} (| \phi \rangle \langle \phi |))$. Because of the $\bm{\alpha}$-dependence of $c(\bm{\alpha})$ in \eqref{fullrecov}, it follows that mixed states are generally only completely recoverable, if the error channel is trace-preserving. A mixed state $\rho = \sum_{\bm{\alpha}_j} p_j | \psi(\bm{\alpha}_j)\rangle \langle \psi(\bm{\alpha}_j )|$ takes the form
\begin{equation}
\mathcal{R} ( \mathcal{E} (\rho) ) =  \sum_{\bm{\alpha}_j} c(\bm{\alpha}_j) p_j | \psi(\bm{\alpha}_j)\rangle \langle \psi(\bm{\alpha}_j )|
\end{equation}
after recovery.
 Pure states, however, are recoverable for trace-preserving and non-trace-preserving error channels.

\section{Existence of a basis}\label{app:basis}
Let us prove that there exists a countable basis $\{ | \psi (\bm{\alpha}_j) \rangle \}$ for the  Hilbert space $\mathcal{C}$. Since $\mathcal{C}$ is separable, there exists a countable orthonormal basis $\{ | n \rangle \}$. Each basis vector can be rewritten as 
\begin{equation}
| n \rangle = \sum_{j} a_{n,i} | \psi (\bm{\alpha}_{n,i}) \rangle 
\end{equation}
because of $\mathcal{C} = \overline{\text{span}} \{ | \psi ( \bm{\alpha } \rangle \}$.
For an element  $| \phi \rangle \in \mathcal{C}$ we can then write
\begin{equation}
| \phi \rangle = \sum_{n} b_n |n \rangle = \sum_{n} \sum_{i} b_n a_{n,i}  | \psi (\bm{\alpha}_{n,i}) \rangle
\end{equation}
Thus, $\{ | \psi (\bm{\alpha}_{n,i}) \rangle \} \equiv \{ | \psi (\bm{\alpha}_{j}) \rangle \}$ is a countable basis of $\mathcal{C}$. By removing elements of the basis, we can ensure that every finite subset of the basis is linearly independent.
\section{Proof that $U_n$ is well-defined} \label{welldefined}
Because of the  possible overcompletness of the basis, it is not immediately apparent that $U_n$ is a well-defined linear operator. Therefore, consider the operator series $U_n^{(N)}$ defined via
\begin{align}
U_n^{(N)} | \phi \rangle = \sum_{i=0}^{N} a_i U_n ( \bm{\alpha}_i)  | \psi( \bm{\alpha}_i) \rangle .
\label{unitaryseries}
\end{align}
Each particular operator $U_n^{(N)}$ is bounded and well-defined which follows from the linear independence of finite subsets of the basis $\{ | \psi (\bm{\alpha}_i) \rangle \}$, in this case $\{ | \psi(\bm{\alpha}_0) \rangle, | \psi(\bm{\alpha}_1) \rangle, \ldots , | \psi(\bm{\alpha}_N) \rangle \}$.  The series $U_n^{(N)} | \phi \rangle$ converges  for $N \rightarrow \infty$ with the limit $U_n | \phi \rangle = \sum_{i}^{\infty} a_i U_n (\bm{ \alpha}_i) | \psi( \bm{\alpha}_i) \rangle$. According to the uniform boundedness principle \cite{Werner}, $U_n$ which defines the limit of the series, is a well-defined linear operator.

\section{Basis-independence of $U_n$} \label{app:independence}
To show basis independence of $U_n$, we pick a basis $\{ |\psi ( \tilde{\bm{\alpha}}_j) \rangle \}$ different from $\{ |\psi (\bm{\alpha}_j) \rangle \}$. Every finite subset of $\{ |\psi ( \tilde{\bm{\alpha}}_j) \rangle \}$ is linearly independent. With this  basis the states in $\mathcal{C}$ can be represented as $| \phi \rangle = \sum_i \tilde{a}_i | \psi (\tilde{\bm{\alpha}}_i)\rangle $ and we define the operator
\begin{align}
\tilde{U}_n | \phi \rangle = \sum_{i=0}^{\infty} \tilde{a}_i U_n ( \tilde{\bm{\alpha}}_i)  | \psi( \tilde{\bm{\alpha}}_i) \rangle 
\end{align}
which is well-defined following the same reasoning as before. 
Let us now show that $\tilde{U}_n$ and $U_n$ do not depend on the basis with which they are defined, that is $U_n P= \tilde{U}_n P$ where $P$ is the projector of $\mathcal{C}$.
To prove equivalence, the following expression needs to be $0$ for all $| \phi \rangle \in \mathcal{C}$
\begin{align}
& \| U_m | \phi \rangle - \tilde{U}_n | \phi \rangle \|^2  = \| (U_m - \tilde{U}_n )| \phi  \rangle \|^2  \nonumber \\
 & = \langle \phi | (U_m^{\dag} - \tilde{U}_n^{\dag}) (U_m - \tilde{U}_n) | \phi \rangle \nonumber \\
& = \langle \phi | U_m^{\dag} U_m - \tilde{U}_n^{\dag} U_m - U_m^{\dag} \tilde{U}_n + \tilde{U}_n^{\dag} \tilde{U}_n | \phi \rangle . \label{basisindependence}
\end{align}
To further evaluate \eqref{basisindependence}, we examine 
\begin{align}
\langle \phi | \tilde{U}_n^\dagger U_m | \phi \rangle &=  \sum_{i,j} \tilde{a}_j^* a_i \langle \psi( \tilde{\bm{\alpha}}_j ) | U_n^\dagger ( \tilde{\bm{\alpha}}_j) U_m ( \bm{\alpha_i}) | \psi( \bm{\alpha}_i ) \rangle \nonumber  \\
&=  \sum_{i,j} \tilde{a}_j^* a_i  \langle \psi( \tilde{\bm{\alpha}}_j) | \psi( \bm{\alpha}_i \rangle \Gamma_{nm} \nonumber \\
&= \langle \phi | \phi \rangle \Gamma_{nm} \Rightarrow P \tilde{U}_n^\dagger U_m P = P \Gamma_{nm}
\label{unitaryproof}
\end{align}
where we used \eqref{orthogonalcritapp} and \eqref{polardecompapp} to get from the first to the second line. Similarly one can show $\langle \phi | U_n^\dagger \tilde{U}_m | \phi \rangle = \langle \phi | \phi \rangle \Gamma_{nm} $ and  $\langle \phi | U_n^\dagger U_n | \phi \rangle =\langle \phi | \tilde{U}_m^\dagger \tilde{U}_m | \phi \rangle= \langle \phi | \phi \rangle$. We find 
\begin{align}
 \| U_m | \phi \rangle - \tilde{U}_n | \phi \rangle \|^2  = \langle \phi | \phi \rangle (2-2\Gamma_{nm})  .
\end{align} 
Since $\Gamma_{nn}=1$, we have $U_n P = \tilde{U}_n P$, which means the operator $U_n P$ does not depend on the choice of basis with which it was defined. For $\Gamma_{nm}=1$ we find $U_n P = U_m P$.

\section{A necessary condition}
Let us now find a necessary condition for the existence of a QEC code for a subclass of states.
Suppose we have a set of errors which is correctable by the recovery operation $\mathcal{R}$ with recovery operators $\{ R_j \}$. The error and recovery operators have to satisfy $R_j E_m | \psi (\bm{\alpha}) \rangle  = \lambda_{jm}(\bm{\alpha}) | \psi (\bm{\alpha}) \rangle$.
By requiring the recovery operation to be  trace preserving, that is $\sum_{j} R_j^\dagger R_j = I$, we obtain
\begin{align}
\langle \psi (\bm{\alpha}) | E_n^\dagger  E_m | \psi (\bm{\beta}) \rangle &= \sum_j \langle \psi (\bm{\alpha}) | E_n^\dagger R_j^\dagger R_j E_m | \psi (\bm{\beta}) \rangle \nonumber \\
&= \sum_{j} \lambda^*_{jn}(\bm{\alpha}) \lambda_{jm}(\bm{\beta}) \langle \psi (\bm{\alpha}) | \psi (\bm{\beta}) \rangle \nonumber \\
&= w_{nm} (\bm{\alpha} , \bm{\beta}) \langle \psi (\bm{\alpha}) | \psi (\bm{\beta}) \rangle \label{necessarycondition}
\end{align}
where $w_{nm}(\bm{\alpha}, \bm{\beta}) = w^*_{mn}(\bm{\beta}, \bm{\alpha})$.
The condition \eqref{necessarycondition} is a necessary one for trace preserving recovery operations. However, it is not a sufficient one. We note that in an infinite dimensional Hilbert space there may be recovery operations, that do not satisfy $\sum_{j} R_j^\dagger R_j = I$, but still happen with probability 1. These QEC would not necessarily satisfy \eqref{necessarycondition}.

\section{Approximate case}
\label{sec:approximate}
Let us assume the NLQEC criterion is approximately satisfied, that is 
\begin{multline}
\langle  \psi( \bm{\alpha} ) |  F_n^\dagger F_m  |   \psi( \bm{\beta} ) \rangle \\=  c_n^* (\bm{\alpha}) c_m(\bm{\beta}) \langle \psi( \bm{\alpha}) |  \psi( \bm{\beta})\rangle \Gamma_{nm} +\epsilon_{nm}(\bm{\alpha},\bm{\beta}).
\label{orthogonalcritapp2}
\end{multline}
with $|c_n(\bm{\alpha})^* c_m(\bm{\beta})| \gg |\epsilon_{nm}(\bm{\alpha},\bm{\beta})|$. Following the same logic as in the proof \ref{proof}, the error operators can be decomposed as $F_n P_{\bm{\alpha}} = c_n (\bm{\alpha}) (1 + \epsilon_{nn}(\bm{\alpha}) / |c_n(\bm{\alpha})|^2) U_n (\bm{\alpha})P_{\bm{\alpha}} $. Using \eqref{orthogonalcritapp2}, we find
\begin{multline}
\langle \psi (\bm{\alpha})| U_n^\dag (\bm{\alpha}) U_m (\bm{\beta}) |\psi (\bm{\beta}) \rangle \\\approx \langle \psi (\bm{\alpha})|\psi (\bm{\beta}) \rangle \Gamma_{nm}(1 - \frac{\epsilon_{nn} (\bm{\alpha}, \bm{\alpha}) }{2 |c_{n} (\bm{\alpha})|^2}- \frac{\epsilon_{mm} (\bm{\beta}, \bm{\beta})}{2|c_m(\bm{\beta})|^2} ) \\ 
+\frac{\epsilon_{nm}(\bm{\alpha}, \bm{\beta}) }{ c_n^* (\bm{\alpha}) c_m (\bm{\beta})} .
\end{multline}
  As the terms $|\epsilon/c^2|^2 \ll 1$ are small, the error operators $F_n$ send the alphabet states to almost orthogonal subspace. Thus, we have have approximately the error space structure that is required for quantum error correction.  Errors can be corrected at least in the approximate sense. For example, one could perform a syndrome measurement using orthogonal projectors $P_r = \tilde{U}_r P \tilde{U}_n^\dag$, which have to be chosen such that the corrupted states $F_n |\psi(\bm{\alpha})\rangle$ are close to these subspaces. The alphabet states then collapse with high probability to one of the subspaces. One then applies the operator $\tilde{U}_n^\dagger$ to approximately recover the state.

\section{Coherent states and full photon loss channel}
\label{sec:coherentstates}

We note that the error channel $ {\cal E} =\{  I, a \}   $  is only an approximate model for photon loss, and is more properly described by a trace-preserving channel.  In this case the NLQEC criterion is not strictly satisfied, due to the effect of amplitude damping of the coherent state. However, for small damping the NLQEC can be shown to be approximately satisfied, recovering the same results as above.   This is physically reasonable since the information encoded in the coherent states $ |\alpha \rangle $ is changed due to the error due to damping. For small levels of photon loss, the information in the coherent state is approximately preserved, in agreement with the NLQEC result.  

Let us now consider the more realistic trace-preserving photon loss channel ${\cal E}  =\{ A_0, A_1, \dots \}$ \cite{chuang97,nielsen00} 
\begin{align}
A_k & \equiv \sum_{n=k}^\infty \sqrt{ { n \choose k}} \sqrt{{\gamma}^{n-k} {(1-\gamma)}^{k}} | n-k \rangle \langle n |  \nonumber \\
&= \sqrt{\frac{(1-\gamma)^k}{k !}} \sqrt{\gamma}^{\hat{n}} a^k .
\label{coherentstateerrorfull}
\end{align}
The error operators $A_k$ act on the coherent states like 
\begin{align}
A_k |  \alpha \rangle = e^{- \frac{|\alpha|^2}{2} \epsilon} \sqrt{\frac{\epsilon^k}{k !}}   \alpha^k  
 | \sqrt{\gamma}\alpha \rangle = c_k (\alpha) | \sqrt{\gamma} \alpha \rangle .
\end{align}
where $\epsilon = 1- \gamma $ is the probability of a loss of a single photon.
Unlike the operator $a$, the $A_k$ operators do not posses eigenstates. However, we call the coherent states $| \alpha \rangle$ with $|\alpha| \epsilon \ll 1$ approximate eigenstates of $A_k$, since $| \sqrt{\gamma} \alpha \rangle \approx  | \alpha \rangle $, by which we mean that the two states are close to each in Hilbert space,  which follows from $\| | \sqrt{\gamma} \alpha \rangle - | \alpha \rangle  \| \approx \frac{\epsilon |\alpha|}{2} \ll 1$.  From the continuity of the inner product follows that $\langle  \sqrt{\gamma} \alpha  | \sqrt{\gamma} \beta \rangle  \approx \langle \alpha | \beta \rangle$. Thus, the NLQEC criterion is approximately satisfied for $\epsilon |\alpha| \ll 1$
\begin{align}
V_{nm} (\alpha, \beta)  &=   c_{n}^* (\alpha) c_m (\beta) \langle \sqrt{\gamma} \alpha | \sqrt{\gamma} \beta \rangle \Gamma_{mn} \nonumber \\
& \approx   c_{n}^* (\alpha) c_m (\beta) \langle  \alpha | \beta \rangle \Gamma_{mn}
\end{align}
where $\Gamma_{nm} = 1$ for all $n,m$, which implies that there is only one error space. In this example, the alphabet states are the coherent states that satisfy $\epsilon |\alpha| = \epsilon \langle n \rangle \ll 1$. For the recovery operator, the theory suggests
\begin{align}
R = I 
\end{align}
which recovers the alphabet states in an approximate sense. We note that Fock states with $\epsilon \langle n \rangle \ll 1$ generally do not retain their identity as they turn into mixed states after passing the error channel. Thus, the alphabet states are uniquely protected in comparison to arbitrary superposition states of Fock states $|n\rangle$.

For the cat state example (Example 3) in the main text, similar results follow.  The even subspace $P_0$ is approximately recoverable for $\epsilon |\alpha| \ll 1$, while an approximate recovery for the subspace $P_1$ is possible for $|\alpha| \gg 0$ and $\epsilon |\alpha| \ll 1$, where $\epsilon$ is the probability of loss of a single photon.


\end{document}